# Texture changes during thermal processing of food: experiments and modelling

**Ankita Sinha[1], Atul Bhargav[1,*]**

[1]Department of Mechanical Engineering
Indian Institute of Technology Gandhinagar
Palaj, Gandhinagar, GJ 382355
India

[*]**Corresponding author**
Tel: +91 792 395 2419, +91 814 030 7813
Email: atul.bhargav@iitgn.ac.in  (Atul Bhargav)

# Texture changes during thermal processing of food: experiments and modelling

Abstract

Texture is an important attribute in the quality assessment of processed food products. Young's modulus is an indirect measure of texture. During thermal treatment of hygroscopic foods, parameters such as moisture content significantly affect Young's modulus. However, the sensitivity to these parameters has not yet been quantified in terms of the stress-strain behaviour. We have built an experimentally validated model to address this gap. This paper presents the stress-strain behaviour and its sensitivity towards various parameters. Experiments are conducted with potato samples for stress-strain behaviour, parametric sensitivity analysis, estimation of initial and critical values of moisture content and Young's modulus. We found that the Young's modulus and the ultimate strength vary by as much as 54% and 29% depending on the rate of applied strain, indicating the need for test standards. Further, we propose a model to predict the local Young's moduli as a function of moisture content, and a relationship between these and the effective Young's modulus. While model results agree well for drying, they deviate by as much as 16% from experiments for frying, indicating the necessity of incorporating additional physics. We expect this work to serve as a crucial step towards the physics-based prediction modelling of local and effective Young's moduli during thermal processing of food products.

## 1. Introduction

Texture plays a key role in the gastronomic acceptability of a food product. This texture is developed throughout the cooking process. Various sensory based texture characteristics that are identified and analysed so far include firmness, springiness, crispiness, cohesiveness and gumminess (Akwetey and Knipe, 2012; Chen and Opara, 2013; Costa et al., 2011; Kim et al., 2012; Konopacka and Plocharski, 2004; Stejskal et al., 2011; Taniwaki and Kohyama, 2012; Wang et al., 2007). Various techniques of texture measurement include texture profile analysis, texture indices measurement and texture modelling (Chen and Opara, 2013). Texture profile analysis involves development of characteristic curves that demonstrate the evolution of texture over time for a food sample. Various researchers have incorporated this measurement method for texture analysis (Chaunier et al., 2007; De Roeck et al., 2010; Farris et al., 2008; Greve et al., 2010; Ragni et al., 2010; Sasikala et al., 2011; Sila et al., 2006). One challenge with texture profile analysis is repeatability, owing to a lack of standardization (Chen and Opara, 2013). Texture indices are empirical relations derived based on mechanical and/or acoustic measurements. Commonly used texture indices are firmness index (Cherng and Ouyang, 2003; Shmulevich et al., 2003), sharpness index (Taniwaki et al., 2009), crunchiness index (Nguyen et al., 2010) and mastication based texture index (Iwatani et al., 2011; Taniwaki et al., 2006; Taniwaki and Sakurai, 2008).

Modelling texture for food products is an alternative approach, which involves predicting texture spatially and temporally using equations that capture the underlying physics. Various empirical, semi-empirical, statistical, logistic and physics-based generic models have been proposed thus far (Chen and Opara, 2013). One of the oldest and most common methods for texture modelling involves the study of texture evolution over time, using first and second order kinetic models with Arrhenius-type rate constants (Hindra and Baik, 2006). Several researchers have used first order kinetic models for analysing various texture attributes for a variety of food materials (Bhattacharya, 2010; Blahovec et al., 2011; Chen and Ramaswamy, 2002; Corzo et al., 2006; De Roeck et al., 2010; Farahnaky et al., 2012; Lana et al., 2005; Liu and Scanlon, 2007; Nisha et al., 2006; Schouten et al., 2007, 2010; Tijskens et al., 2007; Troncoso and Pedreschi, 2007; Zhong and Daubert, 2004). A limited work is also available on the use of second-order kinetic model for texture analysis (Moreira et al., 2008). Due to the ease of application, kinetic models have been extensively used in the field of texture analysis. However, one challenge of such an approach lies in the reusability of estimated empirical constants and their limited applicability to situations other than the ones in which they were estimated. Semi-empirical models such as Maxwell model (Andrés et al., 2008; Bhattacharya, 2010; Del Nobile et al., 2007; Herrero and Careche, 2005; Jain et al., 2007) and Gibson-Ashby equation based models (Le-Bail et al., 2009; Zghal et al., 2002) have also been used for texture analysis. Statistical modelling has also been used by researchers to correlate texture behaviour with various operating parameters (Pinheiro and Almeida, 2008; Subedi and Walsh, 2009; Van Dijk et al., 2006a, 2006b; Yu et al., 2011). Finite element method (FEM) and mechanistic modelling have been used in some instances (Dintwa et al., 2011; Guessasma et al., 2011).

Among other texture characteristics, Young's Modulus has been identified as one of the key parameters that is used to describe texture. Krokida et al. (2000) proposed a non-linear viscoelastic model, which includes maximum stress and strain terms as input parameters that are functions of moisture content. This model is used in some later studies to predict the viscoelastic behaviour of potato strips undergoing compression, and to predict the effect of osmotic and air-drying pre-treatments on the behaviour of French fries (Krokida et al., 2001a, 2001b). (Yang, 2001) empirically determined exponential functions to correlate Young's modulus and fracture stress with moisture content. This model was used to predict large deformation in coupled multi-physics porous media model for microwave drying (Gulati et al., 2016).

Recently developed physics-based modelling frameworks combine multiphase porous media models (Halder et al., 2007a, 2007b) with experimental data to predict the change in local and effective Young's modulus as a function of moisture content and temperature (Thussu and Datta, 2011, 2012), thereby giving an insight into the spatial and temporal evolution of texture. However, generalization of this model may be limited by the need for large number of experimental data points. Another major challenge associated with these models is the wide variation in estimated values of Young's modulus even for very similar samples. These variations in the values of Young's Modulus and stress-strain behaviour makes an investigation of the possible causes associated with data disagreements imperative. One explanation is the geographic, seasonal, agricultural and biological reasons for variation. However, it is also possible that some deviations may occur due to lack of standardised measurement techniques. For instance, the dimensions, aspect ratio, applied strain rate vary significantly across the reported experimental literature (Krokida et al., 2001a, 2001b, 2000b) (Krokida et al., 2001a, 2001b, 2000; Yang, 2001). Similarly, in case of deformation modelling, most of the reported work incorporates empirical models for Young's modulus prediction, this might be a source of error in predicted stress-strain behaviour (Gulati et al., 2016). Thus, there is a need to develop a non-empirical generic model for food materials and to account for the effect of operating parameters. This could serve as a foundation for the development of standardised testing techniques that could be adopted globally.

The current work focusses on investigating the parameters that play a significant role in texture determination in hygroscopic food materials. A predictive model is also developed to estimate Young's modulus as a function of moisture content.

## 2. Objectives

The objectives of the paper are a) analysis of stress-strain behaviour for a selected hygroscopic food material and effect of various physical parameters (geometry, dimensions, aspect ratio, strain rate and moisture content) on Young's modulus and peak strength b) experimental estimation of initial and critical values for moisture content and Young's Modulus for selected samples c) developing a generalized model that could predict local and effective Young's modulus of the food material as a function of local and average moisture content respectively. d) establishing a mathematical relationship between Local Young's moduli and effective Young's modulus of the food material and e) experimental validation of the proposed models. As an additional outcome, we also show the importance of standardizing test protocols for food materials, since this affects the Young's modulus values.

## 3. Experiment

### 3.1 Materials and Methods

Potato (*solanum tuberosum*) of Indian variety *Kufri Badshah* was selected as the sample for the study. Two types of thermal treatments viz. drying and deep frying were analysed. The potato bought from the local market were kept in cold storage before experiments. The samples were brought to room temperature, washed, peeled and air dried before cutting. Each measurement of moisture content, Young's Modulus and temperature were done in triplicate sets and the average values were determined. The sample once used for measuring a parameter was discarded and was not used for measuring other parameters.

Drying phenomena were carried out in a preheated convection oven (Nova Instruments, India). Deep frying was done in a 2-litre electrical fryer (Inalsa Professional2 1700 W electric fryer, India). Texture analyses was done using Ta HD plus Texture Analyzer (Stable Microsystems, UK) with a 70 mm compression platen probe for applying force. Since strain rate changes with the change in sample dimensions, and considering the hypothesis that strain rate may have an effect on the stress-strain behaviour of food samples, test samples were selected such that constant strain rates are maintained as and when required.

### 3.2 Estimation of Input parameters

Samples of 14 X 14 X 2 $mm^3$ were cut to estimate initial properties. Moisture content (MC) was measured using gravimetric method. Samples were kept in a forced convection oven maintained at 120 °C for 24 hours, and the moisture content was estimated from weight difference. To measure the initial Young's modulus (IYM), another cut sample was placed in texture analyzer under the previously stated settings and the stress-strain curve was developed from which the Young's modulus (YM) was estimated later.

Critical Young's modulus (CYM) is defined as the Young's modulus at minimum moisture content (also called critical moisture content, CMC). For measuring the CYM, samples of the previously specified dimensions were taken and kept in a forced convection oven maintained at 120 °C for a period of 24 hrs. Then, the samples were brought out of the oven and cooled until the temperature drops to that of the ambient. The moisture and texture analysis is done as described previously. This procedure was repeated thrice to quantify and ensure repeatability. Figure 1 shows a schematic representation of the procedure.

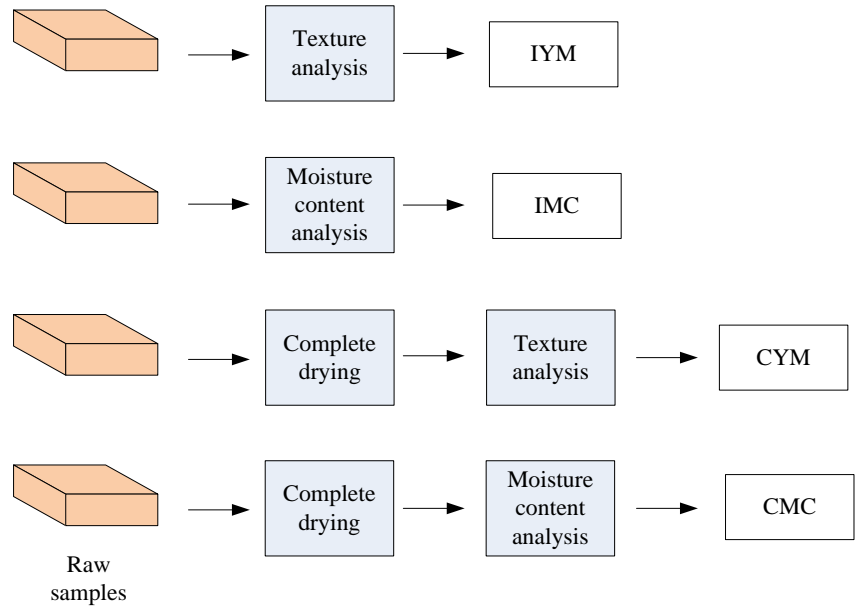

Figure 1: Schematic representation of method for input parameters estimation

## 3.3 Measurement of Local Young's Modulus (LYM)

Potato samples were cut in the square prismatic shape of 14 X 14 mm$^2$ cross section and height 2 mm using a mechanical cutter. The sample height was kept relatively low to ensure near-uniform heating in the sample. This enables one to treat the obtained values of MC and YM as local ones. For the drying experiment, samples were dried in the infrared heater at 115°C for different levels of moisture loss. For deep frying experiments, samples were fried in corn oil at 170°C. for different durations ranging from 0 min to 7 min.

In compression tests, we focus on controlling the strain rate (to 0.01% s$^{-1}$) rather than controlling the test speed (which does not account for sample size variations across tests done by various research groups). These compression tests were carried out in samples with progressive moisture loss, starting with fresh samples and ending with samples with less than 10% moisture, in steps of 10% moisture loss. This gives the variation of LYM with MC. Then, for estimated MC values, the model predicted values of YM is compared with experiments.

## 3.4 Relationship between LYM and EYM

Samples were cut in rectangular prisms 14 X 14 mm$^2$ X 10 cm. For drying experiments, a set of three samples were placed vertically in an infra-red heater, and were dried at a temperature of 115 °C to the desired MC. The drying in the heater occurs due to radiations emitted in the heater. This results in a MC-gradient within the sample. The instrument panel display, however, gives the average MC. After drying, the samples were then taken out and cooled to ambient temperature and then the center 4 cm was considered for further measurements (3 cm lengths on either side were discarded to eliminate end effects). Then, one of the samples was sliced longitudinally into three thin slices and compression testing was performed. YM for each slice was thus determined. These values were treated as the LYM values within the sample. For effective Young's modulus, the other sample with removed end sections was taken and was undergone compression test. The test was done triplicate for each level of moisture content. Similarly, the experiments were performed for samples fried up to different levels of moisture content. Figure 2 gives a schematic representation of the followed procedure.

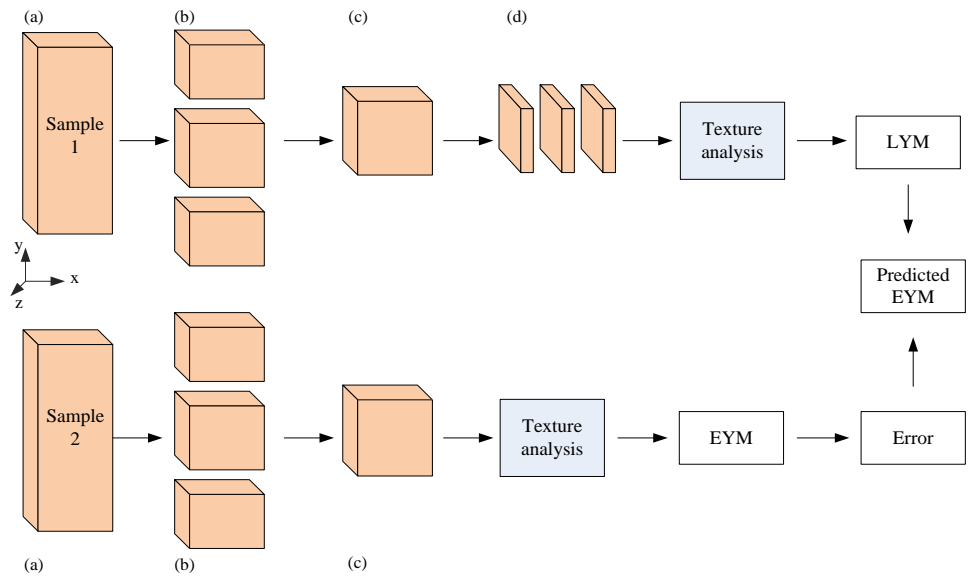

Figure 2: Schematic representation of method for LYM and EYM estimation
Notations: (a) sample (b) sample cut to remove end sections
(c) mid-section with moisture gradient in x direction (d) cut mid-section.

## 4. Modelling

Various studies demonstrate variation in Young's modulus as a function of moisture content (Krokida et al., 2000a; Thussu and Datta, 2011, 2012; Yang, 2001). Due to lack of generalization, the use of these models is restricted. Further, no direct relationship has so far been established between LYM and EYM, so that the EYM could be predicted from LYM values without further experiments. The present work aims to develop a model that could predict LYM and EYM in any food sample. The relationship between LYM and EYM is also derived on the basis of this model.

### 4.1 Model development

Yang (2001) proposed empirical relationships between YM and MC by fitting exponential curves to experimental data. We assume that the variation in Young's modulus follows an exponential trend w.r.t. Moisture content, and a generalized function can be written as:

$$E = \alpha \exp(-\beta M) \tag{1}$$

where $E$ is Young's modulus, $M$ is dry basis moisture content, and $\alpha$ and $\beta$ are fitting parameters. Here, negative sign in the exponential function accounts for the decrease in Young's modulus with increase in moisture content.

The initial and critical values of moisture and Young's modulus for a food sample is a material property and is therefore constant. Thus

at $M = M_0$: $\quad E = E_o \tag{2}$

at $M = M_{cr}$: $\quad E = E_{cr} \tag{3}$

Solving equations (1) (2) and (3), we get:

$$E = E_o \exp\left(\left[\ln\frac{E_{cr}}{E_o}\right]\left[\frac{M_o - M}{M_o - M_{cr}}\right]\right) \tag{4}$$

Equation (4) gives the relationship between moisture content (MC) and Young's modulus (YM).

## 4.2 Relationship between EYM and LYM

Since equation (4) is valid at every spatial position in the sample, at any arbitrary spatial element $k$ with moisture content $M_k$, the YM

$$E_k = E_o \exp\left(\left[\ln\frac{E_{cr}}{E_o}\right]\left[\frac{M_o - M_k}{M_o - M_{cr}}\right]\right) \tag{5}$$

Multiplying for all $k$s

$$\prod_{k=1}^{n} E_k = (E_o)^n \exp\left(\left[\ln\frac{E_{cr}}{E_o}\right]\left[\frac{\sum_{k=1}^{n}(M_o - M_k)}{M_o - M_{cr}}\right]\right) \tag{6}$$

Solving equation (6) and substituting $\frac{\sum_{k=1}^{n} M_k}{n} = M_{av}$:

$$\prod_{k=1}^{n} E_k = (E_o)^n \exp\left(\left[\ln\frac{E_{cr}}{E_o}\right]\left[\frac{M_o - M_{av}}{M_o - M_{cr}}\right]\right)^n \tag{7}$$

$$= \left(E_o \exp\left(\left[\ln\frac{E_{cr}}{E_o}\right]\left[\frac{M_o - M_{av}}{M_o - M_{cr}}\right]\right)\right)^n \tag{8}$$

$$= (E_{eff})^n \tag{9}$$

$$E_{eff} = \sqrt[n]{\prod_{k=1}^{n} E_k} \tag{10}$$

Equation (10) gives the relation between LYM and EYM.

### 4.3 An alternative approach for EYM estimation

Consider a cuboidal sample of length $L$ and cross sectional $A$ having variation in Young's modulus (texture) across its length. The sample can be assumed as a composite member made up of a large number of small elements each having different young's modulus (LYM). Let the total number of small elements be $n$. Therefore, the effective or equivalent Young's modulus (EYM) for composite structure can be calculated as:

For parallel loading:
$$E_p = \frac{E_1 + E_2 + \cdots E_n}{n} \tag{11}$$

For series loading:
$$\frac{n}{E_s} = \frac{1}{E_1} + \frac{1}{E_2} + \cdots \frac{1}{E_n} \tag{12}$$

Equations (11) and (12) give the upper and lower bounds respectively of the EYM

Now, considering that composite is made up of only two members:

$$E_p = \frac{E_1 + E_2}{2} \tag{13}$$

$$\frac{2}{E_s} = \frac{1}{E_1} + \frac{1}{E_2} \tag{14}$$

$$E_s = \frac{2 E_1 E_2}{E_1 + E_2} \tag{15}$$

i.e.
$$E_1 E_2 = E_s E_P \tag{16}$$

Since obtained relations for $E_s$ and $E_p$ correspond to mathematical expressions for arithmetic and harmonic mean which is related to geometric mean as:

$$GM^n = AM * HM \tag{17}$$

Thus,
$$E_g^2 = E_1 E_2 \tag{18}$$

Also, since AM and HM correspond to the upper and lower bound of the possible values, whereas geometric mean is an intermediate approximation.

Now, instead of a 1-D variation in Young's modulus, consider a 3-D structure with varying Young's modulus in each direction. Loading (force application) in such a case would be a combination of both series and parallel loading. Thus, in such a case the better approximation of the effective Young's modulus is an intermediate approximation of series and parallel loading. As already stated, the geometric mean suffices for this criterion, thus effective Young's Modulus can be approximated as $E_g$.

Thus, generalizing equation (18) for n number of elements arranged in a matrix fashion (3-D variation), we get:

$$E_g^n = (E_1 E_2 \dots E_n) \tag{19}$$

Rewriting, $E_g$ as $E_{eff}$ and rearranging, we get:

$$E_{eff} = \sqrt[n]{(E_1 E_2 \dots E_n} \tag{20}$$

$$E_{eff} = \sqrt[n]{\prod_{k=1}^{n} E_k} \tag{21}$$

Equation (21) gives the same expression as derived in equation (10) and thus further justifies the modelling approach.

## 5. Results and Discussion

Results have been arranged in the following order: stress-strain behaviour of raw potato samples under compression testing is shown first; effect of changes in geometry and loading conditions (strain rate) on peak strength (PS) and YM is then presented along with statistical analysis; further, experimental values for initial and critical Young's Modulus is obtained, followed by experimental estimation of LYM at various MC. Experimental results are then compared with model-predicted values. Finally, predicted EYM is also validated with experiments.

### 5.1 Stress strain Behaviour

The stress-strain curve for raw potato sample obtained during compression test on a texture analyser illustrates the behaviour of food materials under loading conditions (Figure 3).

The linear region corresponds to elastic region where Hooke's law is valid. The first peak gives us the ultimate strength of the sample and area under elastic region gives us the resilience. The slope of the elastic region gives the required YM for the sample.

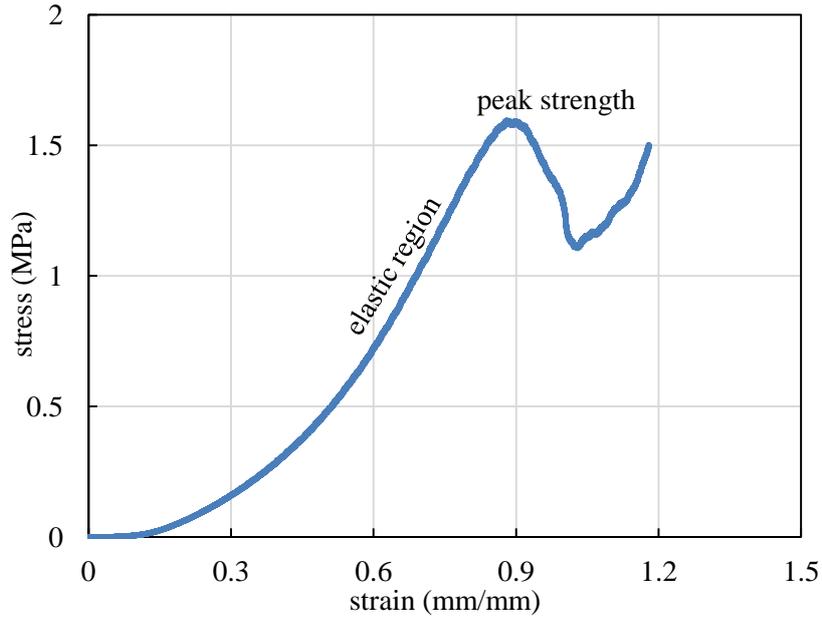

Figure 3: A typical stress-strain curve of potato samples used in this study clearly shows the linear region (where YM is estimated) and the peak strength point.

## 5.2 Effect of experimental procedure on Young's modulus(YM) and peak strength (PS)

The wide variation in YM values for potato samples reported in the literature could be attributed to two factors: 1) sample-to-sample variation (due to the effect of geography, climate & weather, agricultral practices, and storage conditions). 2) variation in test procedures. While the effect of the first factor is largely unavoidable, it is important that the effect of the second factor be investigated and quantified. In order to investigate this, various combinations of physical parameters were selected for compression testing. The effect of change in sample's shape, geometry, moisture content and applied strain rate on YM and PS estimates is investigated; we demonstrate that although the size, shape or aspect ratio may not significantly affect YM and PS estimates, the strain rate has a statistically significant effect on these properties.

Figure 4 shows the obtained experimental values of Young's Modulus and Peak strength during compression testing of potato samples of different geometries, aspact ratio and dimensions. Table 1 describes the various cases considered. Comparison of test case 1, 2 and 3 shows that there is no signifcant effect of aspect ratio on the PS and YM. However, analysis was not done for large variations in aspect ratios. Comparison with case 4 indicates that geometry does not affect the results significantly (statistical analysis is done by performing one way ANOVA (Analysis of Variance) for a single factor using Analysis Toolpak embedded in Microsoft Excel Professional Plus 2013) with a confidence interval of 95% ($\alpha = 0.05$). In both cases, the obtained $p$-value is greater than $\alpha$, and therefore we may conclude that there is no effect of geometry and aspect ratio on YM and PS obtained. Analyses results are shown in Table 2 and Table 3 respectively.

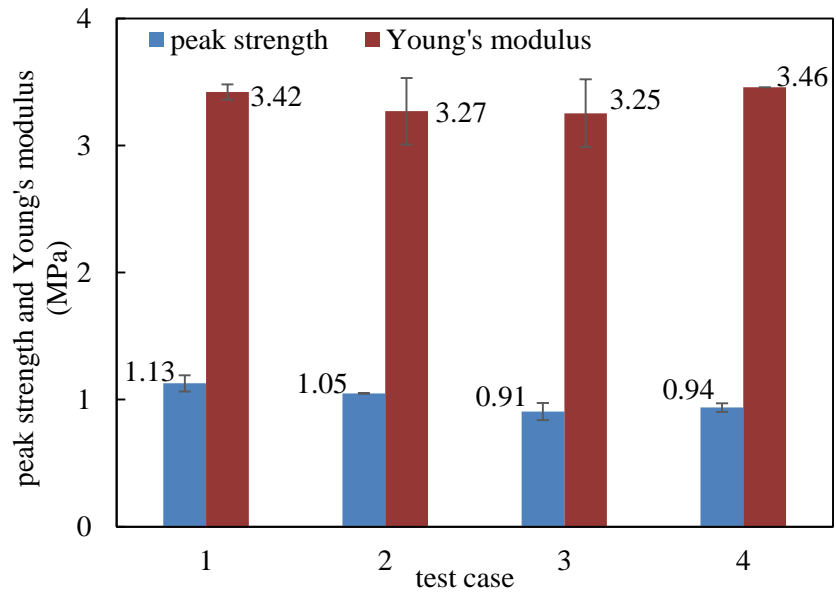

Figure 4: Geometry and aspect ratio do not significantly affect values of YM and PS (see also Table 1).

Table 1: Size and shape details of test cases 1 through 4 (to test for variation in YM and PS values with size, aspect ratio and geometry)

| Test Case | 1 | 2 | 3 | 4 |
|---|---|---|---|---|
| Shape | Cuboid | Cuboid | Cuboid | Cylinder |
| Cross section (mm X mm) | 14 X 14 | 14 X 14 | 7 X 7 | 25 (dia) |
| Height (mm) | 14 | 28 | 14 | 28 |
| Aspect Ratio ($a/l$) | 1:1 | 1:2 | 1:2 | ~1:1 |

Table 2: Analysis of variance (ANOVA) shows that aspect ratio or geometry do not (statistically) affect values of YM

| ANOVA | | | | | | |
|---|---|---|---|---|---|---|
| Source of Variation | SS | df | MS | F | p-value | $F_{critical}$ |
| Between Groups | 0.093 | 3 | 0.031 | 0.239 | 0.867 | 4.066 |
| Within Groups | 1.031 | 8 | 0.129 | | | |
| Total | 1.123 | 11 | | | | |

Table 3: ANOVA shows that aspect ratio or geometry do not (statistically) affect values of PS

| ANOVA | | | | | | |
|---|---|---|---|---|---|---|
| Source of Variation | SS | df | MS | F | p-value | $F_{critical}$ |
| Between Groups | 0.112 | 3 | 0.037 | 2.828 | 0.107 | 4.066 |
| Within Groups | 0.106 | 8 | 0.013 | | | |
| Total | 0.218 | 11 | | | | |

Next, the effect of strain rate on YM and PS was investigated (Figure 5). A similar analysis (as was carried out for the effect of shape and aspect ratio) yielded a statistically significant effect of strain rate on the estimated YM (up to 54% variation) and PS values (up to 29% variation) (Figure 5, Table 4 and Table 5). This indicates the strong need for standardization of test procedures that would reduce type 2 variation in the reported values of YM and PS across the literature.

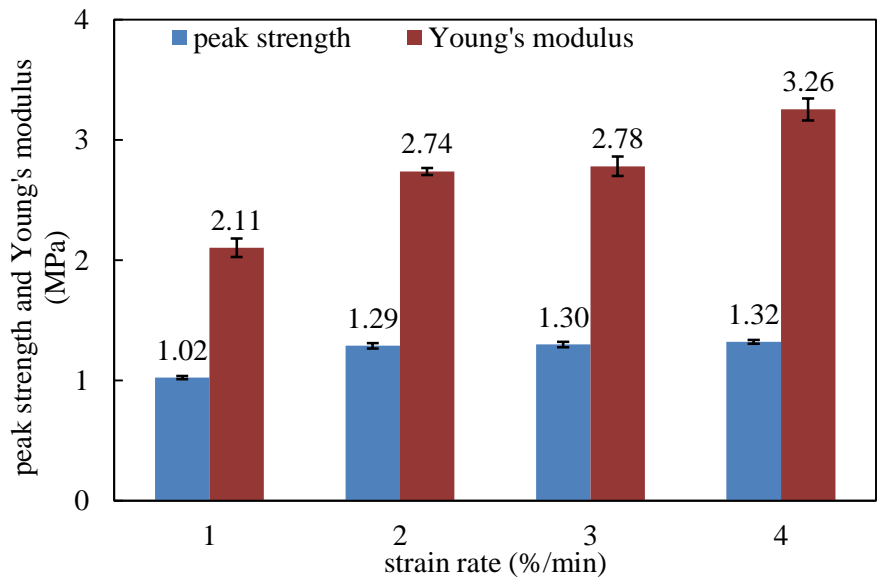

Figure 5: Strain rate affects estimates of YM and PS, indicating the need for standardization of tests for food materials.

Table 4: YM is affected by the strain rate

| ANOVA | | | | | | |
|---|---|---|---|---|---|---|
| Source of Variation | SS | df | MS | F | *p*-value | $F_{critical}$ |
| Between Groups | 3.651 | 3 | 1.217 | 43.763 | $1.79\times10^{-8}$ | 3.160 |
| Within Groups | 0.501 | 18 | 0.028 | | | |
| Total | 4.151 | 21 | | | | |

Table 5: PS is affected by strain rate

| ANOVA | | | | | | |
|---|---|---|---|---|---|---|
| Source of Variation | SS | df | MS | F | p-value | $F_{critical}$ |
| Between Groups | 0.339 | 3 | 0.113 | 56.426 | $2.36 \times 10^{-9}$ | 3.160 |
| Within Groups | 0.036 | 18 | 0.002 | | | |
| Total | 0.375 | 21 | | | | |

### 5.3 Initial and critical values for MC and YM

Using the methods explained in section 3.2, the initial and critical values of moisture content and Young's modulus were determined for potato samples (Table 6).

Table 6: Obtained initial and critical values for moisture content and Young's Modulus

| parameter | value |
|---|---|
| $M_{in}$(db, kg/kg) | 3.73 |
| $M_{cr}$(db, kg/kg) | 0.23 |
| $E_{in}$ (MPa) | 2.87 |
| $E_{cr}$ (MPa) | 6.75 |

### 5.4 Young's modulus of the food material as a function of moisture content

Figure 6 and Figure 7 show the modelling and experimental results for variation in LYM with change in MC for potato samples for drying and deep-frying cases respectively. While there is good agreement for the drying process (figure 6), model predictions deviate significantly for the deep-frying (figure 7).

Starch gelatinization and concomitant softening of the food material, which is not modelled here, is the probable cause of deviation between model predictions and experiments for the deep-frying process. We have developed an improved model that captures effect of gelatinization on texture development, and this model is the subject of discussion in a subsequent paper.

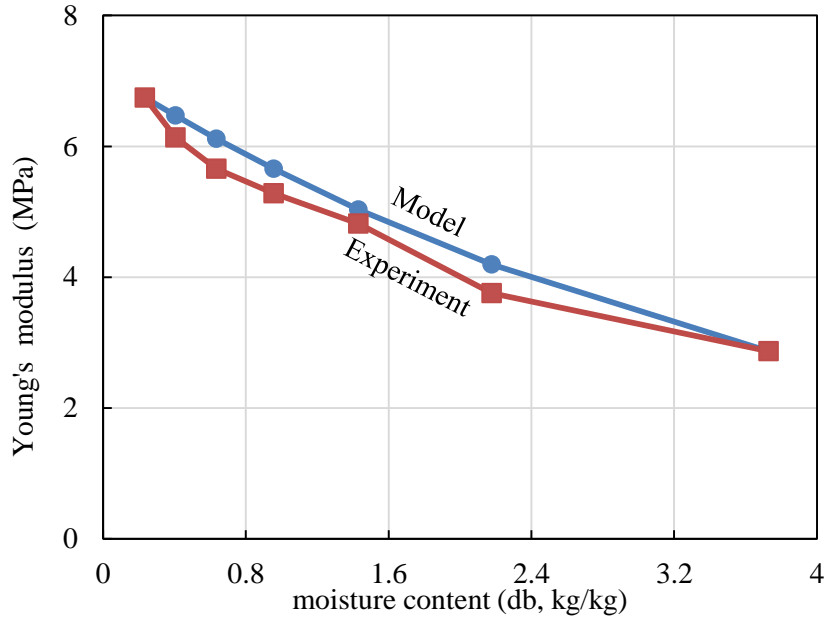

Figure 6: Prediction of variation in local Young's modulus with moisture content during drying shows agreement with experiments.

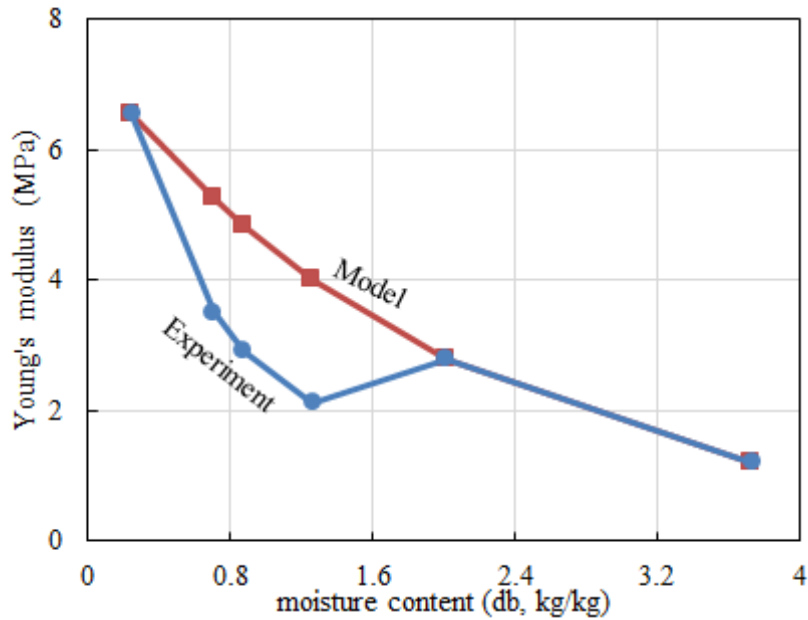

Figure 7: Variation in local Young's modulus with moisture content during deep frying. Model deviates from the experimental results.

## 5.5 Relationship between Local Young's modulus (LYM) and effective Young's modulus (EYM)

As explained in section 3.4, local and effective Young's moduli are measured experimentally. Substituting the experimentally obtained local values in equation (10), EYM is predicted and compared with experimental effective Young's modulus (Table 7 and Table 8). While there is good agreement for partially dried samples at $T = 115$ °C, model deviates for deep-frying at $T = 170$ °C. This deviation (of about 16%) at higher temperatures is also likely caused because gelatinization-based softening is not accounted for in the model

Table 7: A comparison of experimental observations and model predictions shows good agreement for drying of potato samples at 115 °C

| element | LYM/ EYM (MPa) |
|---|---|
| $E_1$ | 3.59 |
| $E_2$ | 2.30 |
| $E_3$ | 1.54 |
| $E_{eff,exp}$ | 2.39 |
| $E_{eff,pred}$ | 2.33 |
| error | 2.44 % |

Table 8: A comparison of experimental observations and model predictions shows significant deviation for deep-fried potato samples

| element | LYM/ EYM (MPa) |
|---|---|
| $E_1$ | 5.57 |
| $E_2$ | 6.25 |
| $E_3$ | 6.07 |
| $E_4$ | 5.93 |
| $E_{eff,exp}$ | 7.12 |
| $E_{eff,pred}$ | 5.95 |
| error | 16.44 % |

## 6. Conclusions

In this paper, Young's modulus, an indicated measure of food texture, is measured and modelled as a function of moisture content. Experimental stress-strain curves are analysed for different measurement conditions, and it is found that strain rate affects YM values by as much as 54%, and PS values by as much as 29%. Therefore, need of standardisation of texture measurement procedures is established through experiments conducted in this work. Further, a model that relates YM with MC is developed and validated with experiments. Model predictions agree with experiments for drying but deviate by as much as 16% for frying. This deviation is postulated to be because the model does not account for starch gelatinization, which causes temporary softening of the food material before it begins to harden again.

The model can be used in multi-physics modelling for predicting transient texture development and shape deformation during thermal food processing.

## Nomenclature

| | |
|---|---|
| $E$ | Young's modulus (MPa) |
| $M$ | moisture content (db, kg/kg) |

Abbreviations
| | |
|---|---|
| YM | Young's modulus |
| PS | peak Strength |
| MC | moisture content |
| IMC | initial moisture content |
| CMC | critical moisture content |
| IYM | initial Young's modulus |
| CYM | critical Young's modulus |
| LYM | local Young's modulus |
| EYM | effective Young's modulus |
| ANOVA | analysis of variance |
| AM | arithmetic mean |
| GM | geometric mean |
| db | dry basis |

subscripts:
| | |
|---|---|
| o | initial |
| cr | critical |
| av | average |
| eff | effective |
| exp | experimental value |
| pred | predicted value |
| g | geometric |

ANOVA:
| | |
|---|---|
| ss | sum of squares |
| df | degree of freedom |
| MS | mean of squares |
| F | estimated F statistics value: |
| $p$-value | estimated probability corresponding to F |
| $F_{critical}$ | F value corresponding to $p$-value $=\alpha=0.05$ |

## Funding

Ankita Sinha receives stipend support from the Ministry of Human Resource Development (MHRD), Govt. of India. This research did not receive any specific grant from funding agencies in the commercial, public or not-for-profit sectors.

## Acknowledgments

The authors are thankful to College of Food Processing Technology and Bio-Energy, Anand Agricultural University for providing access to their experimental facilities for this research work.